\documentclass[12pt]{article}

\begin{document}
\pagestyle{empty}
\begin{flushright}
{OUT--4102--89\\
21 July 2002}
\end{flushright}
\vspace*{3mm}
\begin{center}
{\Large \bf
Bjorken unpolarized and polarized sum rules:\\[3pt]
comparative analysis of large-$N_F$ expansions}

\vspace{0.1cm}

{\bf D.J. Broadhurst$^{a}$ and A.L. Kataev$^{b}$}\\
\vspace{0.1cm}
$^{a}$
Physics and Astronomy Department, Open University,\\
Milton Keynes, MK7 6AA, UK\\

$^{b}$
Institute for Nuclear Research of the Academy of Sciences of
Russia,\\ 117312, Moscow, Russia\\
\end{center}
\begin{center}
{\bf ABSTRACT}
\end{center}
\noindent
Analytical all-orders results are presented for
the one-renormalon-chain contributions to the
Bjorken unpolarized sum rule for the $F_1$
structure function of $\nu{\rm N}$ deep-inelastic scattering
in the large-$N_F$ limit.
The feasibility of estimating higher order
perturbative QCD corrections, by the process
of naive nonabelianization (NNA),
is studied, in anticipation of measurement of
this sum rule at a Neutrino Factory.
A comparison is made with similar estimates obtained for the
Bjorken polarized sum rule.
Application of the NNA procedure to
correlators of quark vector and scalar currents,
in the euclidean region, is compared with
recent analytical results for the $O(\alpha_s^4N_F^2)$ terms.
\vspace*{0.1cm}
\noindent
\\[3mm]
\vfill\eject

\setcounter{page}{1}
\pagestyle{plain}

\section{Introduction}

The differential cross-section for neutrino (anti-neutrino) nucleon
deep-inelastic scattering (DIS) is parametrized by
three structure functions in the familiar formula
\begin{equation}
\frac{d^2\sigma}{dxdy}=\frac{G_F^2 M_{\rm N} E_{\nu}}
{\pi(1+Q^2/M_{\rm W}^2)^2}\bigg[
y^2xF_1+\bigg(1-y-\frac{M_{\rm N}xy}{2E_{\nu}}\bigg)F_2
\pm \bigg(y-\frac{1}{2}y^2\bigg)xF_3\bigg]
\end{equation}
with $0\leq x\leq 1$, $y=E_{\rm had}/E_{\nu}$,
$0\leq y\leq 1/(1+xM_{\rm N}/2E_{\nu})$, and a $+(-)$ sign
applying to neutrino (anti-neutrino) beams.

Analyses of data from previous $\nu{\rm N}$ DIS experiments
concentrated on the extraction of $F_2$ and $xF_3$
(see e.g.\ the works of Ref.\cite{experiments})
and the ratio $R=(1+4M_{\rm N}^2x/Q^2)F_2/(2xF_1)$.
(For a recent extraction in $\nu{\rm N}$ DIS, see Ref.\cite{R}.)
However, there are already at least two attempts
to obtain direct information about $F_1^{\nu{\rm N}}$ \cite{F1}.
Such efforts are interesting from several
points of view. They allow comparison with models for the $O(1/Q^2)$
contributions to both $F_1$ and $xF_3$,
obtained within the framework of an infrared renormalon approach,
which predicts similar $x$-dependence of these two
power corrections \cite{Dasgupta:1996hh}.
A second point of interest is comparison with
QCD sum rule estimates \cite{Braun:1987ty}
for the twist-4 matrix element introduced
in Ref.\cite{Shuryak:1982kj} which gives an $O(1/Q^2)$
correction to the unpolarized Bjorken sum rule (Bjunp SR)
\begin{eqnarray}
C_{\rm Bjunp}(Q^2)
&=& \int_0^1 dx\bigg[F_1^{\nu p}(x,Q^2)-F_1^{\overline{\nu}p}(x,Q^2)\bigg]
\nonumber\\
&=& \int_0^1dx\bigg[F_1^{\nu p}(x,Q^2)-F_1^{\nu n}(x,Q^2)\bigg]
\end{eqnarray}
with a corresponding twist-4 term
for the Bjorken polarized sum rule
estimated in Ref.\cite{Balitsky:1989jb}.

In this note, we present analytical all-orders results
for the one-renormalon-chain contributions to $C_{\rm Bjunp}$,
obtained within the large-$N_F$ expansion, and then apply
the simplistic procedure of naive nonabelianization (NNA),
proposed in Ref.\cite{Broadhurst:1994se}. A comparison
is made with the polarized Bjorken sum rule, and with
the application of NNA to vector and scalar correlators.
We hope that these considerations will encourage
theoretical study of the Bjorken unpolarized sum rule,
whose experimental study may be enabled by $\nu{\rm N}$ DIS data
from a future Neutrino Factory, operating in a region of
medium $Q^2$, with $N_F=3,4$ active flavours.
(For active consideration of such a facility,
see Ref.\cite{Mangano:2001mj}.)

\section{Large-$N_F$ series}

Here we study radiative corrections to Bjunp SR induced in the large-$N_F$
limit. Using methods developed for the calculations in
Refs.\cite{Broadhurst:1993si,Broadhurst:1993ru,Broadhurst:2000yc},
the large-$N_F$ limit of the perturbative part of Eq.(2) was obtained
with the result:
\begin{eqnarray}
C_{\rm Bjunp}&=& 1+\frac{C_F}{T_FN_F}\sum_{n=1}^{\infty}U_n
\bigg(T_FN_F\overline{a}_s\bigg)^n+O(1/N_F^2)\nonumber\\
U_n &= &\lim_{\delta\rightarrow 0}
\bigg(-\frac{4}{3}\frac{{\rm d}}{{\rm d}\delta}\bigg)^{n-1}
U(\delta)
\end{eqnarray}
where $\overline{a}_s$=$\alpha_s(\mu^2=Q^2)/4\pi$ is the coupling
in the $\overline{\rm MS}$-scheme, $C_F=4/3$, $T_F=1/2$ and
\begin{equation}
U(\delta)=-\frac{2\exp(5\delta/3)}{(1-\delta)(1-\delta^2/4)}\,.
\end{equation}
This expression produces the following large-$N_F$ $\overline{\rm MS}$-scheme
series
for the Bjunp SR
\begin{eqnarray}
\sum_{n<10}U_nx^n&=&-2x+\frac{64}{9}x^2-\frac{2480}{81}x^3+\frac{113920}{729}
x^4-\frac{6195968}{6561}x^5+\frac{395898880}{59049}x^6\nonumber\\
&&-\frac{29418752000}{531441}x^7
+\frac{2510236057600}{4782969}x^8-\frac{242876551331840}{43046721}x^9\,.
\end{eqnarray}
It should be noted that in this new result the order $x^2$ and
$x^3$-terms
are in agreement with the corresponding parts of the
total expression for the $\overline{a}_s^2$-correction
to the Bjunp SR \cite{Gorishny:1983gs} and its
$\overline{a}_s^3$-term, calculated
in the $\overline{\rm MS}$-scheme in Ref.\cite{Larin:1990zw}.
Notice also that the series of Eq.(5) has sign-alternating behaviour with
factorially increasing coefficients.
This pattern is explained by the
dominant role of the renormalon at $\delta=1$,
which is generated by a single chain of quark-loop
insertions into the corresponding one-loop QCD diagrams
for this characteristic of $\nu{\rm N}$ DIS.

It is interesting to compare the above
expressions to the analogous ones, obtained in Ref.
\cite{Broadhurst:1993ru}, for the coefficient function
$C_{\rm Bjp}(Q^2)$ in the
Bjorken polarized sum rule (Bjp SR)
\begin{equation}
\int_0^1 dx
\bigg[g_1^{p}(x,Q^2)-g_1^{n}(x,Q^2)\bigg]=
\frac13\left|\frac{g_A}{g_V}\right|
C_{\rm Bjp}(Q^2)\,.
\end{equation}
The large $N_F$-expression
for the Bjp SR
can be obtained
from the following equation \cite{Broadhurst:1993ru}
\begin{eqnarray}
C_{\rm Bjp}&=&1+\frac{C_F}{T_FN_F}\sum_{n=1}^{\infty}
K_n\bigg(T_FN_F\overline{a}_s\bigg)^n
+O(1/N_F^2)\nonumber\\
K_n&=&\lim_{\delta\rightarrow0}
\bigg(-\frac{4}{3}\frac{{\rm d}}{{\rm d}\delta}\bigg)^{n-1}
{K}(\delta)
\end{eqnarray}
where
\begin{equation}
{K}(\delta)\,=\,\left(\frac{3+\delta}{2(1+\delta)}\right)
{U}(\delta)\,=\,-\frac{(3+\delta)\exp(5\delta/3)}
{(1-\delta^2)(1-\delta^2/4)}\,.
\end{equation}
The all-orders large-$N_F$ result of Eq.(8)
was given in Ref.\cite{Broadhurst:1993ru}. The expansion
up to order $O(\alpha_s^9N_F^8)$
was also given in Ref.\cite{Broadhurst:1993ru} in the
$\overline{\rm MS}$-scheme and reads:
\begin{eqnarray}
\sum_{n<10}K_nx^n&=&-3x+8x^2-\frac{920}{27}x^3+\frac{38720}{243}x^4-
\frac{238976}{243}x^5+\frac{130862080}{19683}x^6
\nonumber\\
&&-\frac{10038092800}{177147}x^7
+\frac{274593587200}{531441}x^8-\frac{82519099473920}{14348907}x^9\,.
\end{eqnarray}
As in the case of Eq.(5), this series
is dominated by the renormalon at $\delta=1$,
which has the same reside in each sum rule. (Note that the renormalon
in $K(\delta)$ at $\delta=-1$ is suppressed by a factor
$\frac12\exp(-10/3)=0.018$, relative to the dominant renormalon at
$\delta=1$.)

In the next section we apply naive nonabelianization
to the large-$N_F$ series for both unpolarized and polarized Bjorken
sum rules, in an attempt to estimate higher-order
perturbative coefficients.

\section{Application of NNA to euclidean sum rules}

By naive nonabelianization (NNA) we simply mean the
 substitution \cite{Broadhurst:1994se}
$N_F\rightarrow-\frac{3}{2}\beta_0=N_F-\frac{33}{2}$
in the leading terms of the large-$N_F$ expansion.
The hope is that terms of lower order in $N_F$ may be roughly
estimated by assuming that they follow the leading terms with
weights generated by the one-loop term of the QCD beta function,
$\beta_0=\frac{11}{3}C_A-\frac{4}{3}T_FN_F$.
In a variety of cases (see e.g.\ Refs.\cite
{Broadhurst:1994se,Broadhurst:2000yc,Beneke:1994qe,Lovett-Turner:1995ti,Kataev:2001kk})
this simplistic procedure
gives reasonable estimates of known higher-order
perturbative coefficients in different physical quantities
(see Ref.\cite{Beneke:1999ui} for a review).
As a rule, the signs of contributions from
sets of diagrams with fewer quark
loops are correctly predicted and often the actual magnitudes
of coefficients of lower powers of $N_F$ are within
a factor of 2 of the NNA estimates.
In this section we apply this procedure to the
Bjorken unpolarized and polarized sum rules, where $Q^2>0$
is, of course, in the euclidean region.

We expand the perturbative contributions to
Eq.(2) in powers of $\alpha_s/\pi=4\overline{a}_s$
\begin{equation}
C_{\rm Bjunp}=1+\sum_{n\geq 1}d_n \bigg(\frac{\alpha_s}{\pi}\bigg)^n
\end{equation}
with known results $d_1=-\frac23$ and
\begin{eqnarray}
d_2 &=& - 3.8333 + 0.29630N_F\nonumber\\
d_3 &=& - 36.155 + 6.3313N_F - 0.15947N_F^2
\end{eqnarray}
from Refs.\cite{Gorishny:1983gs,Larin:1990zw},
in the $\overline{\rm MS}$-scheme.
Taking the input of Eq.(5), naive nonabelianization yields
\begin{eqnarray}
d_2^{\rm NNA} &=& - 4.8889 + 0.29630N_F\nonumber\\
d_3^{\rm NNA} &=& - 43.414 + 5.2623N_F - 0.15947N_F^2\nonumber\\
d_4^{\rm NNA} &=& - 457.02 + 83.094N_F - 5.0360N_F^2 + 0.10174N_F^3
\end{eqnarray}
obtained from the exact results for the terms with
highest powers of $N_F$.
One can observe reasonable agreement between the estimates
and exact results for the coefficients
of $N_F^0$ in $d_2$ and of $N_F^1$ in $d_3$. Moreover,
even the estimate, $-43.414$, of the $N_F^0$ term
in $d_3$ has the correct sign
and a magnitude close to the known value, $-36.155$.

We now perform a similar analysis for the
Bjp SR perturbative series $C_{\rm Bjp}(Q^2)=1+\sum_{n\ge1}
\overline{d}_n(\alpha_s/\pi)^n$.
The exact known results are $\overline{d}_1=-1$ and
\begin{eqnarray}
\overline{d}_2 &=& - 4.5833 + 0.33333N_F\nonumber\\
\overline{d}_3 &=& - 41.440 + 7.6073N_F - 0.17747N_F^2\,,
\end{eqnarray}
obtained at $O(\alpha_s^2)$ in Ref.\cite{Gorishny:1985xm}
and at $O(\alpha_s^3)$ in Ref.\cite{Larin:tj}.
The NNA estimates are
\begin{eqnarray}
\overline{d}_2^{\rm NNA} &=& - 5.5 + 0.33333N_F\nonumber\\
\overline{d}_3^{\rm NNA} &=& - 48.316 + 5.8565N_F - 0.17747N_F^2
\nonumber\\
\overline{d}_4^{\rm NNA} &=& - 466.00 + 84.728N_F - 5.1350N_F^2
+ 0.10374N_F^3
\end{eqnarray}
with $\overline{d}_2$ and $\overline{d}_3$
in tolerable agreement with the exact results of Eq.(13).

More ambitious estimates are presented in Tables 1 and 2, where
we have evaluated Eq.(12) and Eq.(14) for $N_F=3,4,5$ active flavours
and compared these estimates with exact results from Eq.(11) and Eq.(13),
or with estimates in Ref.\cite{Kataev:1995vh} at $O(\alpha_s^4)$,
obtained by a method of effective charges \cite{Grunberg:1982fw}.
(The latter are in fair
agreement with Pad\'e estimates in Ref.\cite{Samuel:1995jc}.)

\begin{center}
\begin{tabular}{||r|c|c|c|c||} \hline
${\rm order}$ & $N_F=3$ & $N_F=4$
&$N_F=5$
\\ \hline\hline
$(\frac{\alpha_s}{\pi})^2$
 (NNA) & -4 & -3.70
& -3.40
\\ \hline
$(\frac{\alpha_s}{\pi})^2$
 (exact)& -2.94 & -2.65
&-2.35
\\ \hline\hline
$(\frac{\alpha_s}{\pi})^3$
 (NNA) & -29.1 &-24.9
& -21.1
\\ \hline
$(\frac{\alpha_s}{\pi})^3$
 (exact)& -18.6 & -13.4
& -8.5
\\ \hline\hline
$(\frac{\alpha_s}{\pi})^4$
(NNA) &-250 & -199
&-155
\\ \hline
$(\frac{\alpha_s}{\pi})^4$ (~\cite{Kataev:1995vh}~) &
-133 & -76 & -29
\\ \hline\hline
\end{tabular}
\end{center}
{{\bf Table 1.} The $N_F$-dependence of the NNA expressions
for Bjunp SR and their comparison with the results
of explicit calculations and $O(\alpha_s^4)$
estimates of Ref.\cite{Kataev:1995vh}.}

\begin{center}
\begin{tabular}{||r|c|c|c|c||} \hline
${\rm order}$ & $N_F=3$ & $N_F=4$
& $N_F=5$
\\ \hline\hline
$(\frac{\alpha_s}{\pi})^2$
 (NNA) &-4.5 & -4.17
& -3.83
\\ \hline
$(\frac{\alpha_s}{\pi})^2$
 (exact)& -3.58 & -3.25
& -2.92
\\ \hline\hline
$(\frac{\alpha_s}{\pi})^3$
 (NNA) & -32.3 & -27.7
&-23.5
\\ \hline
$(\frac{\alpha_s}{\pi})^3$
 (exact)& -20.2 & -13.8
& -7.8
\\ \hline\hline
$(\frac{\alpha_s}{\pi})^4$
(NNA) &-255 & -203
& -158
\\ \hline
$(\frac{\alpha_s}{\pi})^4$ (~\cite{Kataev:1995vh}~) &
-130 & -68& -18
\\ \hline\hline
\end{tabular}
\end{center}
{{\bf Table 2.} The $N_F$-dependence of the NNA expressions
for Bjp SR and their comparison with the results
of explicit calculations and
$O(\alpha_s^4)$ estimates of Ref.\cite{Kataev:1995vh}.}

Here one is pushing NNA rather hard, since there are
substantial cancellations between powers of $N_F$ with
alternating signs, which are most pronounced at $N_F=5$.
Nevertheless, there is agreement to within a factor of 2
for $N_F=3,4$ at order $\alpha_s^2$ and $\alpha_s^3$.

The similarities between Tables
1 and 2 are rather striking:
the NNA estimates, the known values, and the effective-charges
estimates in the two sum rules follow very similar patterns.
Indeed this was observed in the renormalization-group invariant
analysis of known results in Ref.\cite{Gardi:1998rf},
where this similarity between polarized and unpolarized
sum rules appeared to be somewhat mysterious. From our point of
view there is a simple-minded explanation: the residues
of the dominant infrared renormalon, at $\delta=1$, in the
Borel transforms of Eq.(4) and Eq.(8) are {\em identical}.
Thus one may attribute the close similarities in the
full perturbative structures to the rough success of NNA,
which assumes that the overall trends are driven by this renormalon.

However, in the concluding section we remind the reader
of two important effects that suggest caution in relying
on estimates obtained from a single chain of quark loops.

\section{Application of NNA to euclidean correlators}

To sharpen our understanding of the limitations of NNA,
we reconsider analyses of the vector \cite{Lovett-Turner:1995ti}
and scalar \cite{Broadhurst:2000yc} correlators, of light-quark
currents, following the recent and impressive calculations
of the $O(\alpha_s^4N_F^2)$ terms reported in
Ref.\cite{Baikov:2001aa}.

In the vector channel we study the Adler function
\begin{equation}
D(Q^2)=Q^2\int_0^{\infty}
\frac{R(s)}{(s+Q^2)^2}ds
=3\sum_F Q_F^2\bigg[1+\sum_{n\geq
1}d^{\rm V}_n\bigg(\frac{\alpha_s}{\pi}\bigg)^n\bigg]
\end{equation}
where $R(s)$ is the well-known $e^+e^-$ ratio, $Q_F$ are the quarks
charges, and the known perturbative results, at large $Q^2>0$
in the euclidean region, are $d^{\rm V}_1=1$ and
\begin{eqnarray}
d^{\rm V}_2 &=& 1.9857 - 0.11530N_F\nonumber\\
d^{\rm V}_3 &=& 18.243 - 4.2158N_F + 0.086207N_F^2\nonumber\\
d^{\rm V}_4 &=& d^{\rm V}_{4,0}+d^{\rm V}_{4,1}N_F
+1.8753N_F^2 - 0.010093N_F^3
\end{eqnarray}
taking $d_2^{\rm V}$ from Ref.\cite{Chetyrkin:bj} and
$d_3^{\rm V}$ from Ref.\cite{Gorishny:1990vf},
with neglect of the term in the vector correlator
with two quark loops and a three-gluon intermediate state.
The $O(\alpha_s^4N_F^3)$ term in $d^{\rm V}_4$
was obtained in Ref.\cite{Beneke:1992ch},
and the $O(\alpha_s^4N_F^2)$ term was recently published
in Ref.\cite{Baikov:2001aa}.
The corresponding NNA estimates are
\begin{eqnarray}
d^{\rm V,NNA}_2 &=& 1.9024 - 0.11530N_F\nonumber\\
d^{\rm V,NNA}_3 &=& 23.470 - 2.8448N_F + 0.086207N_F^2\nonumber\\
d^{\rm V,NNA}_4 &=& 45.338 - 8.2433N_F + 0.49959N_F^2 - 0.010093N_F^3\,.
\end{eqnarray}
While the pattern of agreement of signs and rough magnitudes
is comparable to that for the sum rules it is notable that
the $O(N_F)$ term in $d^{\rm V}_3$ exceeds the NNA estimate
by a factor $1.5$ and the $O(N_F^2)$ term in $d^{\rm V}_4$
exceeds the estimate by a factor $1.8753/0.49959\approx3.8$.

Proceeding to the scalar correlator, again in the euclidean region,
we study the following analogue of Eq.(15)
\begin{equation}
D^{\rm S}(Q^2)=Q^2\int_0^{\infty}\frac{R^{\rm S}(s)}{(s+Q^2)^2}ds
=3\left[m(Q^2)\right]^2
\bigg[1+\sum_{n\geq
1}d^{\rm S}_n\bigg(\frac{\alpha_s}{\pi}\bigg)^n\bigg]
\end{equation}
with $d_1^{\rm S}=\frac{17}3$ and
\begin{eqnarray}
d_2^{\rm S} &=& 51.567 - 1.9070N_F\nonumber\\
d_3^{\rm S} &=& 648.71 - 63.742N_F + 0.92913N_F^2\nonumber\\
d_4^{\rm S} &=& d_{4,0}^{\rm S}+
d_{4,1}^{\rm S}N_F+ 54.783N_F^2 - 0.45374N_F^3
\end{eqnarray}
taking $d_2^{\rm S}$ from
Ref.\cite{Gorishny:1990zu},
$d_3^{\rm S}$ from
Ref.\cite{Chetyrkin:1996sr},
the $O(\alpha_s^4N_F^3)$ term in $d^{\rm S}_4$
from Ref.\cite{Broadhurst:2000yc},
and the $O(\alpha_s^4N_F^2)$ term
from Ref.\cite{Baikov:2001aa}.
We choose this comparator so as to make contact
with the analyses of Refs.\cite{Chetyrkin:1997wm,Baikov:2001aa},
notwithstanding the comment made in Ref.\cite{Broadhurst:2000yc}
that the necessity of a second subtraction in the dispersion
relation for the scalar correlator makes such a construct
infrared unsafe. The corresponding NNA estimates are
\begin{eqnarray}
d_2^{\rm S,NNA} &=& 31.465 - 1.9070N_F\nonumber\\
d_3^{\rm S,NNA} &=& 252.96 - 30.661N_F + 0.92913N_F^2\nonumber\\
d_4^{\rm S,NNA} &=& 2038.3 - 370.60N_F + 22.460N_F^2 - 0.45374N_F^3
\end{eqnarray}
again with a fair pattern of agreement in signs and magnitudes.
The $O(N_F)$ term in $d^{\rm S}_3$ exceeds the NNA estimate
by a factor $2.1$ and the $O(N_F^2)$ term in $d^{\rm S}_4$
exceeds the estimate by a factor of
$54.783/22.460\approx2.4$

Two remarks are in order.
First, one notes that the absolute sizes of the
radiative corrections in the scalar channel
are much larger than those in the vector channel.
This may be regarded as a success
for the NNA estimator, which attributes this trend
to the inadequately subtracted
dispersion relation of Eq.(18), taken as the object of
study in Refs.\cite{Chetyrkin:1997wm,Baikov:2001aa}.
The failure to make a second subtraction in the scalar
channel results \cite{Broadhurst:2000yc}
in an infrared renormalon at $\delta=1$,
whereas no such renormalon can appear in Adler's
correctly subtracted dispersion relation of Eq.(15).
Moreover, this disparity between the scalar and vector
channels is already apparent in the $O(\alpha_s)$
diagrams, into which quark loops are inserted, where the
coefficient $d_1^{\rm S}=\frac{17}{3}$ is almost 6 times larger
than $d_1^{\rm V}=1$.

As in the previous section, this renormalon analysis
has value only if the known radiative
corrections follow the pattern suggested by NNA.
Fortunately this is again the case, in the euclidean
region. We remark that the
very large $O(\alpha_s^4N_F^2)$ coefficient $d_{4,2}^{\rm S}=
54.783$ is better approximated by NNA than is the far smaller
coefficient $d_{4,2}^{\rm V}=1.8753$.
It thus appears that NNA is a useful indicator
of significant euclidean radiative corrections.

The second important observation
is that a very different picture emerges if one chooses
the comparator $R^{\rm S}(s)$, which is the
discontinuity of the scalar correlator across the cut in the
minkowskian (timelike) region $-Q^2=s>0$.
The modest success of NNA in the euclidean region
does not ensure comparable success in the minkowskian region,
because of the numerically large terms involving
powers of $\pi^2$ and coefficients of the beta function,
$\beta(\alpha_s)$,
and the mass anomalous dimension, $\gamma_m(\alpha_s)$,
noted in \cite{Chetyrkin:1997wm}. These easily computed
effects of analytic continuation do {\em not\/} naively
nonabelianize. Moreover, the infrared renormalon at $\delta=1$ in
Eq.(18) is absent \cite{Broadhurst:2000yc} from
the scalar imaginary part $R^{\rm S}(s)$.
Accordingly, NNA behaves poorly in this comparator.
For example, it was noted in
Ref.\cite{Baikov:2001aa} that
the exact $O(\alpha_s^4N_F^2)$ term in $R^{\rm S}(s)$
exceeds that obtained by applying NNA
to the exact $O(\alpha_s^4N_F^3)$ term of
Ref.\cite{Broadhurst:2000yc}
by a factor of $9.6848/1.0128\approx10$.

Thus we learn
from the analytical calculations of
Refs.\cite{Broadhurst:2000yc,Baikov:2001aa}
that NNA estimates for the scalar
correlator perform better in the euclidean
region, where we have already noted tolerable
agreement with exact results on sum rules.

\section{Conclusions}
In summary:
\begin{enumerate}
\item The new all-orders results for the leading terms in
the large-$N_F$ expansion of the perturbative contributions
to the Bjorken unpolarized sum rule, when naively
nonabelianized, offer an explanation of the observation
of Ref.\cite{Gardi:1998rf} that the actual radiative corrections
in this sum rule closely follow those of the Bjorken polarized
sum rule. Within the simplistic, but here relatively successful,
framework of NNA, we attribute this parallelism to the equality
of the residues of the dominant infrared renormalon, at $\delta=1$, in the
Borel transforms of Eq.(4) and Eq.(8).
\item The pattern of NNA estimates for the vector and scalar
correlators, in the euclidean region, also suggests a leading
role for infrared renormalons. In this case, it helps one to understand
why the actual scalar radiative corrections are so much
larger than those in the vector case. To attribute this
to NNA, it is necessary for the procedure to have reasonable
success in each channel. In fact we find that it is slightly
more successful in the scalar case, notwithstanding
conspicuous failure after analytic continuation
to the minkowskian region.
\item Since NNA yields correct signs and sensible magnitudes
for all known coefficients of terms with lower powers of $N_F$,
in all four {\em euclidean\/} analyses considered in this work,
we believe that its sign predictions
in Eqs.(12,14,17,20), for unknown $O(\alpha_s^4)$
terms, are probably reliable. In particular, we expect
that in Eqs.(16,19) further dedicated calculation
will yield negative signs for $d_{4,1}^{\rm V,S}$
and positive signs for the $d_{4,0}^{\rm V,S}$ coefficients
that result from purely gluonic radiative corrections.
\item It is important to recognize the intrinsic
limitations of NNA. First we remark that
the $N_F$ dependence of QCD radiative corrections
results not only from insertions of quark loops
in gluon propagators; in addition there are diagrams
with quark loops inserted at gluonic vertices.
It seems unreasonable to expect NNA to mimic contributions such
as that in Fig.1(a) of Ref.\cite{Baikov:2001aa}, where a quark
loop modifies a three-gluon vertex within a second quark loop.
More significantly, perhaps, one first
encounters at $O(\alpha_s^4)$ the effects of
quark loops in different gluon propagators.
As remarked in Ref.\cite{Vainshtein:1994ff},
multiple renormalon chains modify the asymptotic
behaviour at next-to-leading order in $1/N_F$,
notably in the $O(\alpha_s^4N_F^2)$ terms.
Therefore, it would appear more prudent to use
the final row in each of the Tables 1 and 2
as estimates for DIS sum rules. Moreover,
in the vector correlator, the effective-charges
analysis of Ref.\cite{Kataev:1995vh} appears
to be in good agreement with the partial
analytical results of Ref.\cite{Baikov:2001aa}.
\end{enumerate}
\section{Acknowledgements}
This work was first presented at the Quarks-2002 International Seminar
(June 1--7, 2002, Novgorod the Great, Russia).
ALK thanks C.J. Maxwell and A.I. Vainshtein for
comments and acknowledges
support by RFBR Grants N 00-02-17432 and N 02-01-00601.

\end{document}